\documentclass[pdflatex,sn-mathphys-num,iicol]{sn-jnl}

\usepackage{graphicx}%
\usepackage{multirow}%
\usepackage{amsmath,amssymb,amsfonts}%
\usepackage{amsthm}%
\usepackage{mathrsfs}%
\usepackage[title]{appendix}%
\usepackage{xcolor}%
\usepackage{textcomp}%
\usepackage{manyfoot}%
\usepackage{booktabs}%
\usepackage{algorithm}%
\usepackage{algorithmicx}%
\usepackage{algpseudocode}%
\usepackage{listings}%

\theoremstyle{thmstyleone}%
\theoremstyle{thmstyletwo}%

\theoremstyle{thmstylethree}%

\begin{document}

\title[Article Title]{Impact of heavy-tailed synaptic strength distributions on self-sustained activity in networks of spiking neurons.}


\author*[1]{\fnm{Ralf} \sur{T\"onjes}}\email{ralf.toenjes@hu-berlin.de}

\author[2]{\fnm{Chunming} \sur{Zheng}}

\author[3]{\fnm{Wenping} \sur{Cui}}

\author[1,4]{\fnm{Benjamin} \sur{Lindner}}

\affil*[1]{\orgdiv{Institute of Physics}, \orgname{Humboldt University}, \orgaddress{\city{Berlin}, \country{Germany}}}

\affil[2]{\orgdiv{School of Physics and Astronomy}, \orgname{Yunnan University}, \orgaddress{\city{Kunming}, \country{China}}}

\affil[3]{\orgname{Princeton University}, \city{Princeton}, \country{USA}}

\affil[4]{\orgdiv{BCCN}, \orgname{Humboldt University}, \city{Berlin}, \country{Germany}}

\abstract{We analyze states of stationary activity in randomly coupled quadratic integrate-and-fire neurons using stochastic mean-field theory. Specifically, we consider the two cases of Gaussian random coupling and Cauchy random coupling, which are representative of systems with light- or with heavy-tailed synaptic strength distributions. 
For both, Gaussian and Cauchy coupling, bistability between a low activity and a high activity state of self-sustained firing is possible in excitable neurons. 
In the system with Cauchy coupling 
we find analytically a directed percolation threshold, i.e., above a critical value of the synaptic strength, activity percolates through the whole network starting from a few spiking units only. The existence of the directed percolation threshold is in agreement with previous numerical results in the literature for integrate-and-fire neurons with heavy-tailed synaptic strength distribution. However, we have found that the transition can be continuous or discontinuous, depending on the excitatory-inhibitory imbalance in the network. Networks with Gaussian coupling and networks with Cauchy coupling and additional additive noise lack the percolation transition in the thermodynamic limit.}




\maketitle

\section{Introduction}
\label{sec:Introduction}
Neural activity in living biological tissue is a highly complex, stochastic, non-equilibrium and non-linear process. However, due to the large number of constituents, i.e., the interacting spiking neurons, it is possible to use methods from statistical physics to describe such systems in terms of mean-field observables.
In \cite{Bru00} Brunel developed a standard model for homogeneous random networks of spiking neurons. The Brunel model became widely used to numerically reproduce and analytically predict different states of activity in biological neural networks, specifically the experimentally omnipresent asynchronous irregular state \cite{RenDel10}. 

In order to arrive at a feasible mathematical description, Brunel made several simplifying key assumptions, like independence of the different input signals to a neuron and, following from that and the central limit theorem, the Gaussian approximation for the summed inputs arriving at a specific neuron within the network. Convergence of the sum of independent inputs in mean and distribution to a Gaussian requires inputs of bounded variance. However, experimental data show that synaptic strengths in neural networks can be broadly distributed \cite{BarBru07,SonSjo05,BuzMiz14} with possible implications for neural information processing \cite{TerTsu12}.
In this case, a Gaussian description is no longer accurate as the convergence of the input sum is poor in the tails of the distribution and a description in terms of a Gaussian mean-field theory becomes questionable. It has recently been suggested \cite{KusOga20}, that heavy-tailed synaptic weight distributions can change the neural dynamics and, in particular, support a continuous directed percolation transition with order parameter exponent $\beta=1/2$ at the critical point\footnote{The critical exponent $\beta$ describes the leading order scaling of the order parameter close to the bifurcation point or phase transition. It is $\beta=1$ for transcritical bifurcations, $\beta=1/2$ for saddle-node and pitchfork bifurcations and $\beta=1/3$ at the cusp point of a cusp bifurcation. At a transcritical bifurcation two fixed points cross under finite slope and exchange stability, hence $\beta=1$.}.
In marked contrast, such a continuous transition has not been observed in dense networks with light-tailed (e.g. Gaussian) distributions \cite{KusOga20}.  Here, in order to deeper understand the impact of heavy-tailed synaptic distributions 
on the neural activity in recurrent networks of spiking neurons, we compare the mean firing rate in systems of randomly coupled quadratic integrate-and-fire (QIF) neurons with either Gaussian or with Cauchy random synaptic weights.

Excitable neurons in random networks are known to exhibit self-sustained activity \cite{RenMor07}. A nonzero mean of the presynaptic input, impinging on the individual neurons, can shift the working point of excitable neurons into a regime of regular firing. Moreover, the many inputs arriving through the randomly weighted excitatory and inhibitory synaptic connections constitute a dynamic network noise which can also lead to a self-consistent random spiking of the driven neurons, even when the mean is zero.
Stochastic mean-field theory aims at deriving equations for the self-consistent mean values and correlation statistics of the neural spike trains \cite{Lerchner2006,DumWie14,VanLin18,PenVel18,VelLin19}.
Under certain conditions, in random networks a state of self-sustained high activity co-exists with a state of low activity. Either of these states can become unstable upon changes in a system parameter or under external stimulus, and the system may jump discontinuously to the respective other state. First-order, i.e.,discontinuous, phase transitions are of great interest in applications, because of hysteresis. That is, the system persists in the new state and does not directly revert to the original state when the external stimulus or the change of the system parameter is performed in the reverse order. In a second-order phase transition, which in \cite{KusOga20} was predicted for neural networks with heavy-tailed synaptic weights, the order parameter is continuous but has a discontinuous and possibly infinite derivative. Near such a phase transition low-dimensional mean-field theory is no longer an accurate description of the system, because fluctuations and their temporal correlations
can become scale free, i.e., macroscopic. The role of criticality in generating complexity in biological neural networks has been suggested by statistical physicists \cite{StaBak95,HerHop95} and has since also been studied experimentally \cite{BegPle03,PetThi09,SheCla15}.

Bistability and more complicated collective dynamics, such as synchronized oscillations or irregular fluctuations of synchronized activity on slow time-scales have been found and analyzed in random networks of coupled integrate-and-fire neurons with light-tailed positive (excitatory) and negative (inhibitory) coupling 
\cite{Bru00,Ost14,WieBer15,DivSeg22}. All-to-all coupling with independent, identically distributed random weights is an idealization of such random networks, which is often used in statistical, solid-state physics, e.g.,for spin models in disordered media \cite{SheKir75,JanEng10}. 

Here we use this kind of coupling with independent random positive and negative weights as a model for unstructured networks of excitatory and inhibitory spiking neurons focusing on the  non-equilibrium steady states of irregular asynchronous spiking. We explore the influence of the coupling statistics, excitatory-inhibitory balance and additional noise on these steady states in networks of excitable QIF neurons. Using the commonly applied white-noise approximation for the self-generated dynamic noise in the network, we can derive and solve self-consistent mean-field equations for the firing rate. 

For excitable QIF neurons with Cauchy random coupling, we confirm the existence of a direct percolation transition in the absence of external noise. However, a slight imbalance in favor of excitatory synapses provides a positive feedback, rendering the transition discontinuous, and inducing bistability as in the case of Gaussian random coupling. In networks with stronger inhibition bistability persists with Gaussian random synaptic weights whereas the percolation transition in the Cauchy case becomes continuous.

Our paper is organized as follows: In  Sec.\,\ref{sec:TheModel}, we introduce the QIF neuron model and the different coupling network statistics used in our study. We discuss the feedback of the neural activity to the self-generated, dynamic noise in the system and how stochastic mean-field theory (SMFT) can be used to derive closed self-consistent analytical descriptions of the stationary dynamics. The details of the numerical simulation of the dynamics in the full, high-dimensional neuron network model and measures of the neural activity are also presented there. In Sec.\,\ref{Sec:Gauss} we use the known self-consistency theory to inspect the mean firing rate in networks with Gaussian coupling, taking also into account additional Gaussian white noise acting on each element.   
In Sec.\,\ref{Sec:Cauchy} we study the mean-field dynamics in the model with Cauchy random coupling under a novel Cauchy white-noise approximation. Our analytical results are confirmed by 
simulations of recurrent networks of spiking QIF neurons. We conclude in Sec.\,\ref{sec:conclusion} with a summary of the main results. 

\section{Models, measures, and simulations}
\label{sec:TheModel}
In balanced networks of spiking neurons the sequence 
of excitatory and inhibitory inputs from presynaptic neurons constitutes a dynamic noise, which strongly affects the output spike trains of the postsynaptic neurons. We refer to the noise generated by the neurons as dynamic, because its statistics, in particular its strength and mean depend uniquely on the activity itself, i.e., on the average firing rate, and on higher-order statistics of all neurons in the network. The properties of the dynamic noise are part of the dynamics. Stochastic mean-field theory incorporates the statistics of single-neuron spike trains into the characteristics of the noise generated in the network in a self-consistent way. 

Here we limit our analysis to a self-consistent firing-rate which feeds back into the properties of the network noise in a white noise approximation. The exact power spectrum of the spike trains, or the effects of interspike interval variability, e.g.,the coefficient of variation, are not considered. We furthermore disregard the secondary effect of heterogeneity in the stationary firing rates \cite{RoxBru11}. We have found that the stationary network activity and the network noise are well approximated under these simplifying assumptions over a wide range of system parameters.
In the following, we present the models analyzed in this paper.  
\subsection{Models}
We consider QIF neurons with instantaneous, current-based pulse coupling. 
The membrane potentials $v_\ell(t)$, of the $\ell=1\ldots N$ neurons, follow QIF model dynamics in nondimensional time and voltage variables \cite{Izh07}
\begin{equation}\label{Eq:QIF01}
    \dot{v}_\ell = v_\ell^2 + a_0 + z_{\ell}(t).
\end{equation}
Here $a_0$ is the excitation parameter and $z_\ell(t)$ is a nondimensional input current. Regarding the time-independent drift of the uncoupled QIF dynamics, $v_\ell^2+a_0$, we note that
in the excitable regime when $a_0<0$, the membrane potential has a resting point and a saddle, respectively at $\pm\sqrt{-a_0}$, whereas it is spiking tonically for $a_0>0$. 

In addition to the dynamic network noise $z^{NW}_\ell(t)$ from presynaptic neurons, noise $z_\ell^{add}(t)$ from other, independent neural populations or from intrinsic channel fluctuations can also be included as $z_\ell(t)=z_\ell^{NW}(t)+z_\ell^{add}(t)$. Upon crossing of a threshold potential $v_{th}$ at times $t_{m k}$, and reset to a value $v_r$, an action potential, here represented by a $\delta$-kick, is generated and distributed as part of the spike-trains 
\begin{equation}
    x_m(t) = \sum_{k=-\infty}^\infty \delta(t-t_{mk})
\end{equation}
to the inputs $z^{NW}_\ell(t)$ of all postsynaptic neurons
\begin{equation}\label{Eq:zNW}
    z^{NW}_\ell(t) = \sum_{m=1}^N J_{\ell m} x_m(t).
\end{equation}
The coupling coefficients, or synaptic weights, $J_{\ell m}$ are randomly drawn, independently for each connection, either from a Gaussian 
\begin{equation}\label{Eq:GaussianJ}
    J_{\ell m} \sim \mathcal{N}(\mu/N,\sigma^2/N)= \frac{\mu}{N}+ \frac{\sigma}{\sqrt{N}}\mathcal{N}(0,1)
\end{equation}
or a Cauchy distribution
\begin{equation}\label{Eq:CauchyJ}
    J_{\ell m} \sim \mathcal{C}(\mu/N,\sigma/N) = \frac{\mu}{N}+ \frac{\sigma}{N}\mathcal{C}(0,1)
\end{equation}
with Gaussian $$p_\mathcal{N}(J)=\left(2\pi\sigma^2/N\right)^{-\frac{1}{2}}\exp\left(-\frac{(J-\mu/N)^2}{2\sigma^2/N}\right)$$ and Cauchy probability density 
$$p_\mathcal{C}(J) = \frac{1}{\pi} \frac{\sigma/N}{(J-\mu/N)^2+\sigma^2/N^2},$$ respectively.
The scaling in the Gaussian case is such that the standard deviation of the synaptic weights is much larger than its mean value for large network sizes $N$, which corresponds to the often discussed case of a balanced network 
\cite{VanSom96,RenDel10}. 
We refer to $\sigma$, which quantifies the width of the distribution, as coupling {heterogeneity}, and to $\mu$, which quantifies a prevalence for positive or negative synaptic weights, as excitatory-inhibitory imbalance. The scaling of the imbalance and the {heterogeneity} in the synaptic weight distributions with the system size makes $\mu$ and $\sigma$ system-size independent (intensive) parameters, so that the mean-field dynamics in systems of finite size are comparable to the dynamics in a well-defined thermodynamic limit $N\to\infty$. 

Interpreting $v(t)$ as a state variable rather than a physical voltage, it is possible to formally take the simultaneous limits of $v_{th},v_r\to\pm\infty$, because once $v(t)$ exceeds the saddle value it increases according to Eq.\,\eqref{Eq:QIF01} super-exponentially to infinity in a finite time and returns from $v_r=-\infty$ to the neighborhood of the resting potential. For $a_0>0$, there is no resting potential, and the QIF neuron is spiking periodically with period $T=\pi/\sqrt{a_0}$. The QIF model is equivalent to a theta neuron model 
\begin{equation}\label{Eq:Theta01}
    \dot\vartheta_\ell = (1-\cos\vartheta_\ell) + (1+\cos\vartheta_\ell)(a_0+z_{\ell}(t))
\end{equation}
under the non-linear transformation $v = \tan(\vartheta/2)$. The theta neuron dynamics \eqref{Eq:Theta01} has been derived independently from a normal form analysis of a saddle-node bifurcation on an invariant circle (SNIC) \cite{ErmKop86,Erm96}. The neuron generates a spike when $\vartheta$ crosses $\pi$ continuously. This transformation requires the chain rule of differentiation. If the inputs $z_\ell(t)$ have a white-noise component or contain delta pulses, we obtain a so-called multiplicative  noise (see \cite{Gar85,Sok10}) in the transformed dynamics Eq.\,\eqref{Eq:Theta01} that needs an interpretation (see the discussion of this issue for the QIF model driven by Gaussian white noise in Ref.~\cite{LinLon03}). Here we want to interpret Eq.\,\eqref{Eq:Theta01} as equivalent to Eq.\,\eqref{Eq:QIF01}, i.e., 
not in the sense of Ito \cite{Ris84}. 

Networks of QIF and theta neurons allow for an exact, low-dimensional mean-field description under certain conditions. For instance, if all neurons receive the same input signal $z_\ell(t)=z(t)$, Watanabe-Strogatz theory \cite{WatStr94} applies to a finite ensemble of $N$ identical theta neurons and the exact evolution of the system is obtained from the solution of a three dimensional, non-autonomous differential equation \cite{PikRos08}. Furthermore, the Ott-Antonsen ansatz \cite{OttAnt08} can be applied to obtain an exact differential equation for the mean field in the so called thermodynamic limit of $N\to\infty$ for theta neurons and other phase dynamics with quenched heterogeneity or subject to white Cauchy noise \cite{LukBar13,PazMon14,ToePik20}. These analytical results for coupled phase oscillators have successfully been used to derive exact mean-field equations for the mean membrane potential and the mean firing rate in ensembles of QIF neurons subject to non-Gaussian noise \cite{PieCesPik23,GolPer24,CluMon24}.

In this work we assume that the dynamic noise $z_\ell^{NW}(t)$ is approximately independent for different neurons and white (temporally uncorrelated) Gaussian- or Cauchy noise, depending on the distribution of synaptic weights. The noise strength depends through Eq.\,\eqref{Eq:zNW} on the firing rates of all neurons, and the firing rates in turn depend on the noise strength. This input-output relation defines stationary firing rates in the network self-consistently. Self-consistency analysis is commonly applied together with the Gaussian white noise approximation \cite{Bru00,DivSeg22}. For QIF neurons with Cauchy distributed synaptic weights, under the approximation of independent, white Cauchy noise, we modify the mean-field equations derived in \cite{CluMon24} to reflect the feedback of the dynamic network noise, allowing in principle for a full nonlinear analysis of the transient dynamics and the stationary state. 
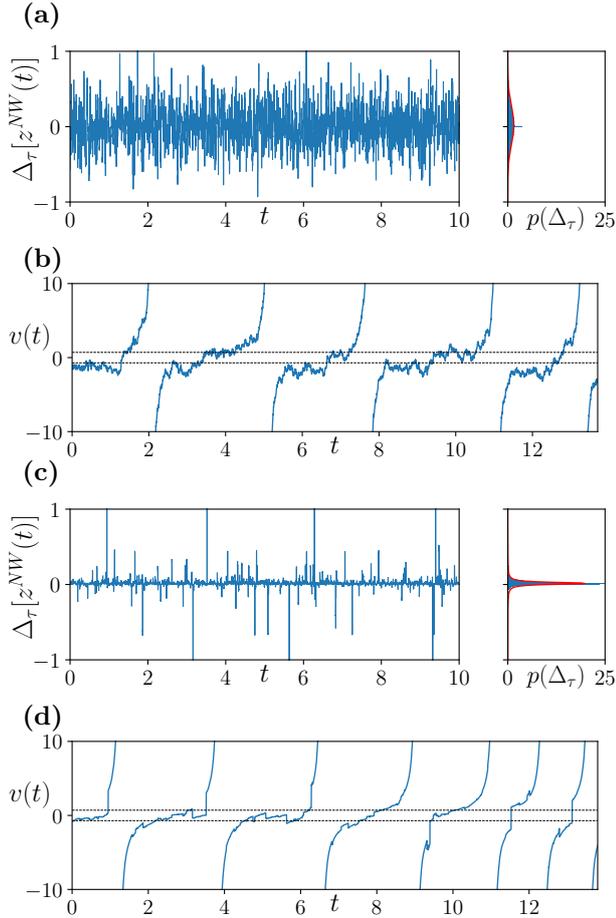
\begin{figure}[t!]
\setlength{\unitlength}{1cm}
\begin{picture}(8.0,3)
\put(-0.5,0){\includegraphics[width=0.515\textwidth]{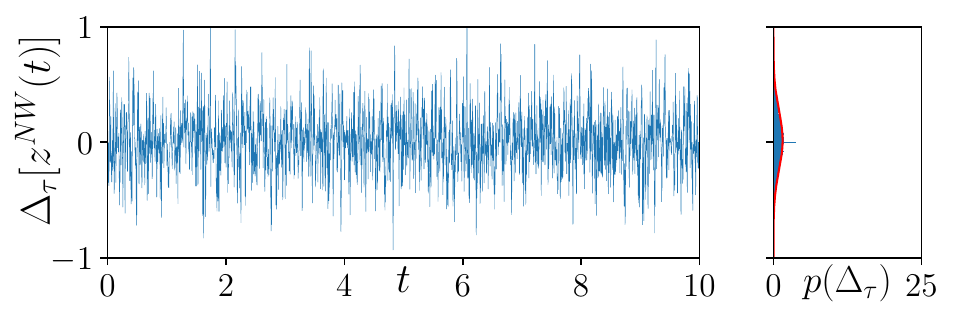}}
\put(-0.2,2.9){\bf (a)}
\end{picture}
\begin{picture}(8,3)
\put(-0.5,0){\includegraphics[width=0.5\textwidth]{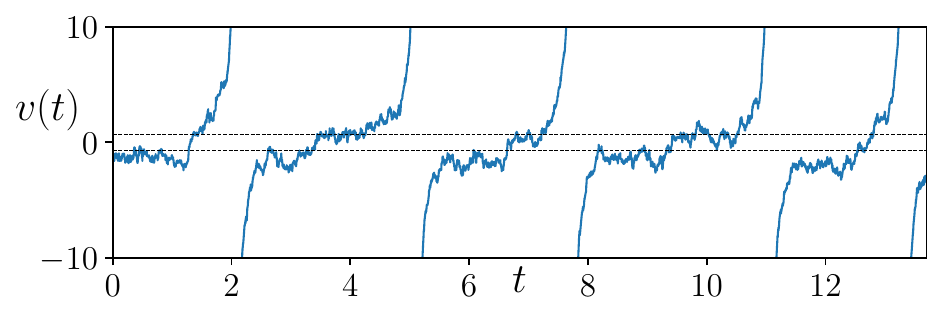}}
\put(-0.2,2.7){\bf (b)}
\end{picture}
\begin{picture}(8.0,3)
\put(-0.5,0){\includegraphics[width=0.515\textwidth]{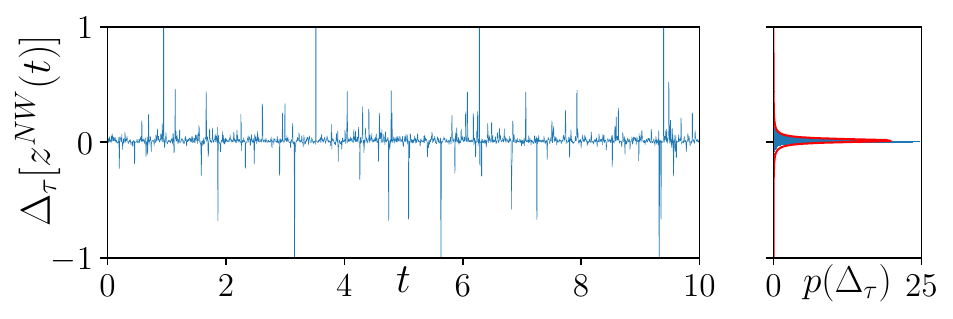}}
\put(-0.2,2.9){\bf (c)}
\end{picture}
\begin{picture}(8,3)
\put(-0.5,0){\includegraphics[width=0.5\textwidth]{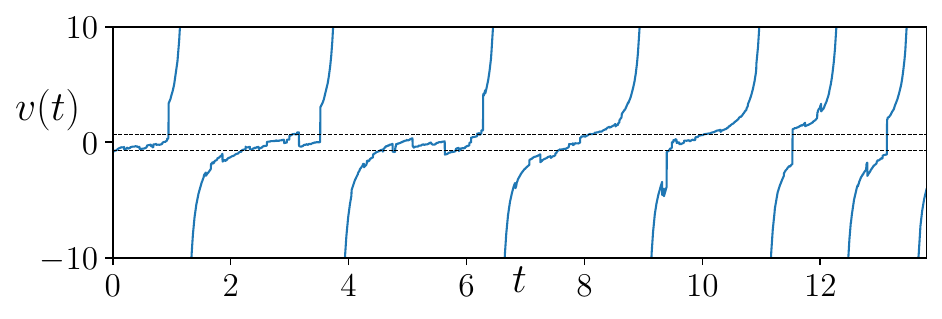}}
\put(-0.2,2.7){\bf (d)}
\end{picture}
\caption{\textbf{Qualitative comparison between the network noise with Gaussian and with Cauchy random synaptic weights.} Short-time, sliding window filtered inputs $\Delta_\tau[z^{NW}_\ell(t)]$ (Eqs.\,\eqref{Eq:GaussDeltaTau} and \eqref{Eq:CauchyDeltaTau}, window width $\tau=10\,\Delta t=0.01$), each for a single excitable neuron ($a_0=-0.5$). Taken from direct simulation of $N=1000$ QIF neurons (a) with Gaussian coupling disorder and (c) with Cauchy coupling disorder. The distribution of these values are shown as horizontal, normalized histograms to the right. Imbalance and {heterogeneity} of the synaptic weights in both examples are $\mu=4$ and $\sigma=4$ and we have not included additional noise to the neuron dynamics (see Figs.\ref{Fig:GaussTheory_vs_Sims}b and \ref{Fig:CauchyTheory_vs_Sims}b). The red curves superimposed on the histograms are the theoretical distributions $\mathcal{N}(\mu r \tau,\sigma^2 r \tau)$ and $\mathcal{C}(\mu r \tau, \sigma r \tau)$ in the Gaussian and in the Cauchy case, using the measured average firing rates in the networks $r=0.41$ and $r=0.36$, respectively. In (b) and (d), below the time series of the linearly filtered noise, we have aligned the corresponding time series of the single neuron membrane potentials $v_\ell(t)$ with jumps clearly visible for Cauchy distributed synaptic strengths. The dashed horizontal lines mark the resting potential and the saddle at $\pm\sqrt{-a_0}$, respectively. The working point of the neuron has shifted in both cases to positive values $a_0+\mu r>0$, i.e.,the self-sustained activity is mean-driven in this example.
}
\label{Fig:NoiseDemo}
\end{figure}

In order to visualize the qualitative difference between the dynamic noise in networks with Gaussian and with Cauchy random synaptic weights we show in Fig.\ref{Fig:NoiseDemo} a short-time sliding window filter of the input $z^{NW}_\ell(t)$ to a single neuron and the corresponding stochastic membrane potential $v_\ell(t)$ in regimes with comparable network activities. The network noise with Cauchy distributed synaptic weights appears to be much sparser. In contrast, noise with Gaussian statistics is much stronger over small time scales, e.g.,the width of the distribution of the integrated noise grows as a square root of the window size, while the distribution of the integrated noise in the case of Cauchy coupling is narrower in the center, with a width that grows linearly in the window size, but has scale-free heavy-tails. In this case, frequent small spikes and rare spikes through strong synapses contribute equally to the stochastic dynamics.
\subsection{Simulations and firing rate estimation}
We perform simulations with networks of $N=1000$ randomly coupled, excitable QIF neurons. The stationary mean firing rates in these simulations are determined as follows. The stationary firing rate $r_\ell$ of a neuron is the average number of spikes per time unit. 
This average is performed by counting the number of spikes, $N_\ell(T)$, in a long time window $T$ for a network with frozen connectivity, and taking ${r}_\ell= \lim_{T\to\infty} N_\ell(T)/T$. 
Furthermore, we average over all neurons to obtain the mean activity of the network, i.e., $r=(1/N)\sum_\ell r_\ell$ 
and also record the average and the variance of the activity over several network realizations.

The mean activity $r$ is  compared to the self-consistent SMFT solutions of the firing rate in the white-noise approximation. For each value of the imbalance parameter $\mu$, we use a single coupling-matrix realization and perform a forward and a backward sweep of the {heterogeneity} $\sigma$ (the values of $J_{\ell m}$ are only adjusted and not redrawn after changing $\sigma$). For each value of $\sigma$ we integrate the system over a transient of $T_\text{trans}=10$ and afterwards measure the firing rates over a longer time window of $T=100$ time units in steps of $\Delta t=0.001$ and determine their ensemble average ${r}$, the network activity, from all neurons. 

To avoid any complication due to the multiplicative nature of the noise in the theta neuron dynamics but still benefit from the continuous (non-diverging) dynamics in the phase variable we carry out a mixed integration scheme switching back and forth between the integration of $v$ and $\vartheta$. In each integration step we first perform the deterministic update of $\vartheta_\ell$ according to Eq.\,\eqref{Eq:Theta01} with $z_\ell=0$ using Euler steps and record the neurons that cross $\pi$. We then add the external noise $\Delta w \sim \sqrt{2 D^{add} \Delta t}\,\mathcal{N}(0,1)$ and the instantaneous shifts $J_{\ell m}$ due to delta kicks from the firing presynaptic neurons $m$ to the membrane voltages $v_\ell$.

At the beginning of a forward sweep in $\sigma$ we reset all neurons to the equilibrium potential, ensuring an initial state of low, externally excited activity, while we start with uniformly random initial phases $\vartheta_\ell$ in the backward sweeps, preparing the system in the basin of attraction of the self-sustained, high activity state. Fluctuations in the activity for a single network realization, and heterogeneity over several network realizations exist due to the relatively small system size. The parameter sweeps are repeated ten times  with different coupling-matrix realizations and averaged. In bistable regimes some averages will fall between the two self-consistent SMFT solutions at parameters where the transition time, due to finite-size fluctuations, is comparable to the measuring period of $T=100$. 

In excitable neurons with Cauchy random coupling and without additional noise, the quiescent state, where all neurons are resting at the fixed point, can become unstable in the mean-field description but remain stable in the forward simulations (see Fig.\ref{Fig:CauchyTheory_vs_Sims}b). In the considered high-dimensional dynamical system, this state is stable against finite perturbations of individual membrane voltages, but as a fixed point of the mean-field dynamics it is unstable against a vanishingly small fraction of neurons that are initially excited, i.e., poised beyond the saddle point.
\begin{figure}[t!]
\setlength{\unitlength}{1cm}
\begin{picture}(3.6,4.0)
\put(-0.21,0.2){\includegraphics[height=0.22\textwidth]{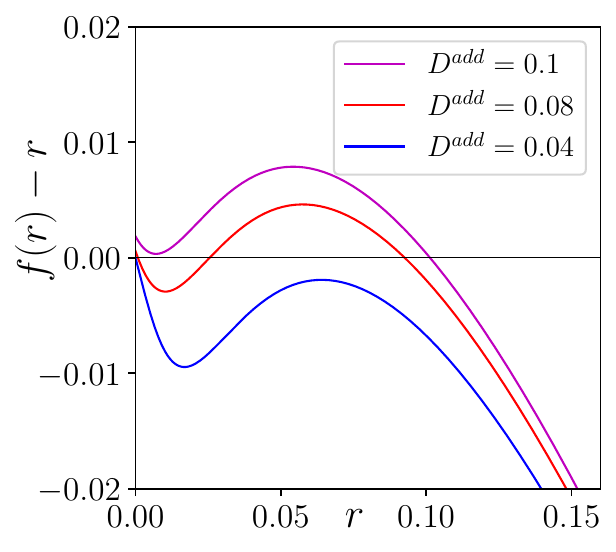}}
\put(0.1,3.8){\bf (a)}
\end{picture}
\begin{picture}(3.7,4.0)
\put(-0.1,0.2){\includegraphics[height=0.22\textwidth]{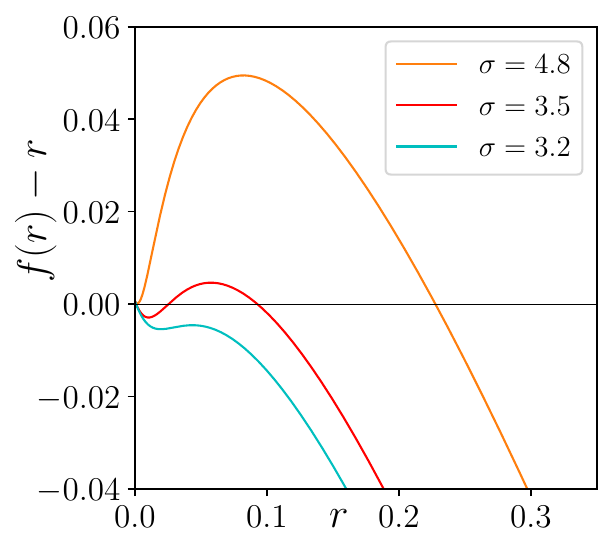}}
\put(0.2,3.8){\bf (b)}
\end{picture}
\caption{\textbf{Input-output relation of pre- and postsynaptic firing rates.} Input-output relations Eq.\eqref{Eq:PrePostMap} for the rates of excitable QIF neurons with $\mu=0$, (a) {heterogeneity} $\sigma=3.5$ and different values of additive noise strength $D^{add}$ and (b) $D^{add}=0.08$ and different values of $\sigma$. Shown is the difference between the post- and the presynaptic firing rates, in the Gaussian white-noise approximation, when the postsynaptic neurons are driven by independent Poissonian presynaptic spike trains with firing rates ${r}^{pre}$. Stable self-consistent rates are located where the difference crosses zero with a negative slope smaller than two. There, a small deviation of ${r}^{pre}$ from that solution will result in a ${r}^{post}$ closer to the solution.
}
\label{Fig:PrePostMap}
\end{figure}
\section{Gaussian coupling}
\label{Sec:Gauss}
\subsection{Gaussian-white-noise approximation}
We first use a Gaussian random coupling matrix Eq.\,\eqref{Eq:GaussianJ} with independent, identically distributed entries. Over a time interval $\tau$ a neuron receives on average $Nr\tau$ input spikes from its presynaptic neighbors, where $r$ is the average firing rate in the system. Because the sum of independent identically distributed Gaussian variables is also Gaussian, with additive mean and variance, the total synaptic input, arriving in a small enough time interval $\tau$ is normally distributed. The time interval $\tau$ should be much smaller than $1/r$ to ensure independence of the spikes but long enough to apply the central limit theorem, i.e., $Nr\tau>1$. Due to our chosen scaling with the system size in Eq.\,\eqref{Eq:GaussianJ} we obtain a system-size independent increment statistics
\begin{equation}\label{Eq:GaussDeltaTau}
    \Delta_\tau[z^{NW}(t)] = \int_0^\tau z^{NW}(t+s)\,ds \sim \mathcal{N}(\mu {r} \tau,  \sigma^2 {r} \tau)
\end{equation}
shown in Fig.\ref{Fig:NoiseDemo}a for a single neuron.
If we furthermore neglect the temporal correlations in the presynaptic spike trains over larger time intervals $\tau$, we can regard Eq.\,\eqref{Eq:GaussDeltaTau} as the integral of a Gaussian white noise, with bias $a^{NW}=\mu r$ and noise intensity $D^{NW}=\sigma^2 r/2$. This is the Gaussian white noise approximation, also known as the diffusion approximation that we employ in the following. 
Deviations from predictions under the diffusion approximation can arise from temporal correlations in the noise \cite{Lerchner2006,DumWie14}, violating the assumption of uncorrelated fluctuations, and from finite synaptic strengths \cite{RicSwa10,DroLin17,GolVolTor24}, violating the Gaussian assumption and making the increments \eqref{Eq:GaussDeltaTau} discontinuous. We stress that in our Gaussian network model the diffusion approximation is accurate in the limit $N\to\infty$, and in our simulations with $N=1000$, but in general it must be checked carefully \cite{GolVolTor24}.

Given that each neuron has the excitation parameter $a_0$ and receives additional white noise $z^{add}(t)$ of intensity $D^{add}$, i.e., $\langle z^{add}_\ell(t)z^{add}_m(t')\rangle = 2D^{add}\delta(t-t')\delta_{\ell m}$, the effective excitation parameter $a=a_0+a^{NW}$ and the effective noise strength each neuron receives $D=D^{add}+D^{NW}$ are modified due to the recurrent input. These two rate-dependent shifts in the system parameters essentially describe the feedback of the self-generated network noise on the dynamics of the neurons.
\subsection{Self-consistent determination of the firing rate}
The firing rates of QIF neurons with excitation parameter $a=a_0+a^{NW}$ and Gaussian white noise input of intensity $D=D^{add}+D^{NW}$ can be calculated as 
\begin{equation}\label{Eq:QIF_rate}
    {r} = \phi(a,D) = \left(\frac{3D}{2}\right)^{\frac{1}{3}} \frac{1}{G(s)}
\end{equation}
with 
\begin{equation}\label{Eq:QIF_s}
    s = -a\left(\frac{12}{D^2}\right)^\frac{1}{3} 
\end{equation}
from a convergent power series \cite{ColSan89,LinLon03}, here presented in rearranged form
\begin{equation}\label{Eq:QIF_Gs}
    G(s) = \sqrt{\frac{\pi}{3}}\sum_{n=0}^\infty \frac{s^n}{n!}\Gamma\left(\frac{2n+1}{6} \right).
\end{equation}
The function $\phi(a+I,D)$ is the single neuron gain function \cite{GerKis02} for an input $I$ at noise level $D$.
The function $G(s)$ was expressed in \cite{DivSeg22} analytically in terms of Bessel functions with fractional index. In the low noise limit $|s|\to\infty$, the series \eqref{Eq:QIF_Gs} converges poorly but the rate \eqref{Eq:QIF_rate} has simple asymptotics \cite{LinLon03}, the Kramer's rate 
\begin{equation}
    {r} \approx \frac{\sqrt{|a|}}{\pi} e^{-\frac{4\sqrt{|a|^3}}{3D}}
\end{equation}
for excitable QIF neurons with $a<0$, $|s|\gg 1$ and
\begin{equation}
    {r} \approx \frac{\sqrt{a}}{\pi}
\end{equation}
for tonically spiking QIF neurons with $a>0$ and $s\gg 1$.
Since $a$ and $D$ depend dynamically on the firing rate ${r}^{pre}$ of the presynaptic neurons, the firing rate ${r}^{post}$ of the post synaptic neuron is a function of $r^{pre}$
\begin{equation}\label{Eq:PrePostMap}
    {r}^{post} = f(r^{pre}) = \phi\left(a_0+\mu{r}^{pre},D^{add}+\sigma^2{r}^{pre}/2\right).\qquad
\end{equation}
Self consistency requires the firing rate in the network in a stationary state of asynchronous, irregular firing to be a fixed point ${r}^{post}={r}^{pre}={r}$ of this map. 
In Fig.\ref{Fig:PrePostMap} we illustrate this mapping by plotting $f(r)-r$ as a function of ${r}$ for excitable QIF neurons with $a_0=-0.5$ and for various system parameters. At self-consistent rates, this difference is zero. 
Stability of the map under iteration requires $f'(r)<1$ which corresponds to a zero crossing of $f(r)-r$ at a negative slope and coincides in all network simulation with the observed stability of the self-consistent stationary firing rates. The self-consistency equation 
\begin{equation}\label{Eq:RateConsistency}
    {r} = \phi\left(a_0+\mu{r},D^{add}+\sigma^2{r}/2\right)
\end{equation}
implicitly defines the mean firing rate, i.e., the stationary activity in the network, as a function of all other system parameters. This equation of state can be solved parametrically, i.e., for $(\sigma(s),{r}(s))$ or $(D^{add}(s),{r}(s))$ (see Appendix \ref{secA1}), or by numerical root-finding and continuation.
\begin{figure}[t!]
\setlength{\unitlength}{1cm}
\begin{picture}(3.6,4.0)
\put(-0.15,0.2){\includegraphics[height=0.235\textwidth]{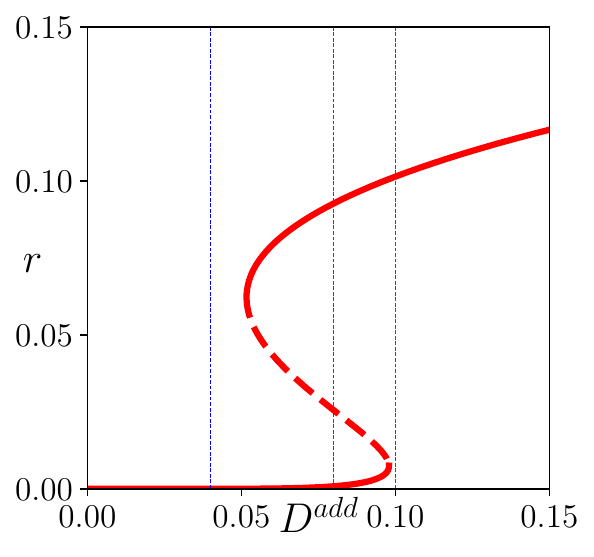}}
\put(0.1,4){\bf (a)}
\end{picture}
\begin{picture}(3.7,4.0)
\put(-0.1,0.2){\includegraphics[height=0.235\textwidth]{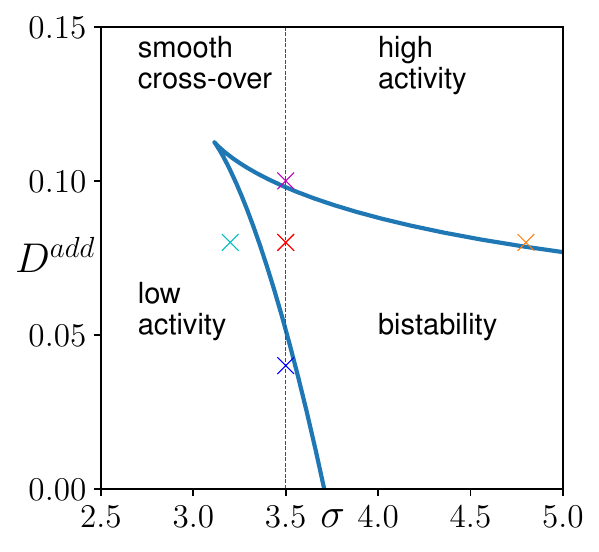}}
\put(0.2,4){\bf (b)}
\end{picture}
\caption{\textbf{Bistability region.} We construct bifurcation diagrams by solving the self-consistency equation Eq.\,\eqref{Eq:RateConsistency} parametrically (see Appendix \ref{secA1}). Shown in (a) is the self-consistent firing rate as a function of the additive noise strength for excitable neurons $a_0=-0.5$, coupling {heterogeneity} $\sigma=3.5$ and zero imbalance $\mu=0$. A stable state of high self-sustained activity co-exists with a stable state of low activity (solid lines), sustained by the additive noise over an interval of bistability. These two solutions are connected by an unstable solution branch (dashed line) and are created or annihilated in saddle-node bifurcations at the knees of the solution curve. With two control parameters, here shown in (b) for the {heterogeneity} $\sigma$ and the additive noise strength $D^{add}$, the surface of self-consistent firing rates has the form of a fold. Bistability exists between the saddle-node bifurcation lines (blue) which connect in a cusp point. The crosses correspond to points in the parameter space where we show the input-output relation for the pre- and postsynaptic firing rate in Fig.\ref{Fig:PrePostMap}. The dotted vertical line marks a cut at $\sigma=3.5$ across the bistability region which we have used in panel (a). The vertical lines in (a) correspond to the crosses in (b) at $\sigma=3.5$ and additive noise strengths $D^{add}\in\{0.04,0.08,0.1\}$, i.e., the input-output relations shown in Fig.\ref{Fig:PrePostMap}a.}
\label{Fig:FoldExplorationGauss}
\end{figure}
\begin{figure}[t!]
\setlength{\unitlength}{1cm}
\begin{picture}(3.6,4)
\put(-0.21,0.2){\includegraphics[height=0.235\textwidth]{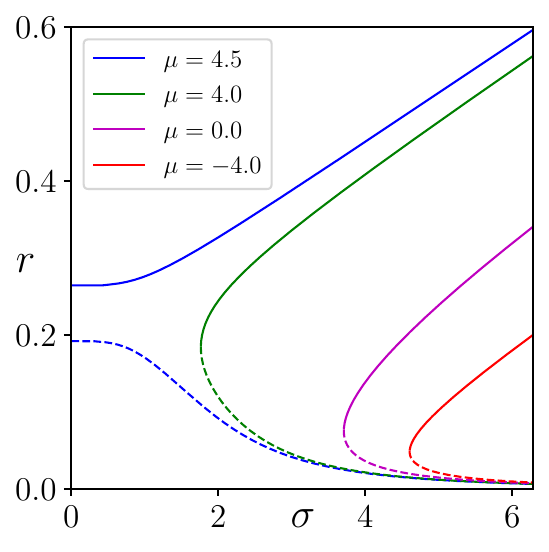}}
\put(0.1,4){\bf (a)}
\end{picture}
\begin{picture}(3.7,4)
\put(-0.1,0.2){\includegraphics[height=0.235\textwidth]{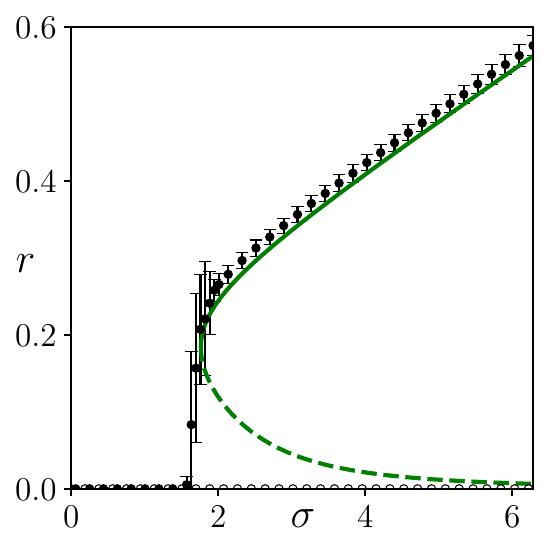}}
\put(0.2,4){\bf (b)}
\end{picture}
\begin{picture}(3.6,4)
\put(-0.21,0.2){\includegraphics[height=0.235\textwidth]{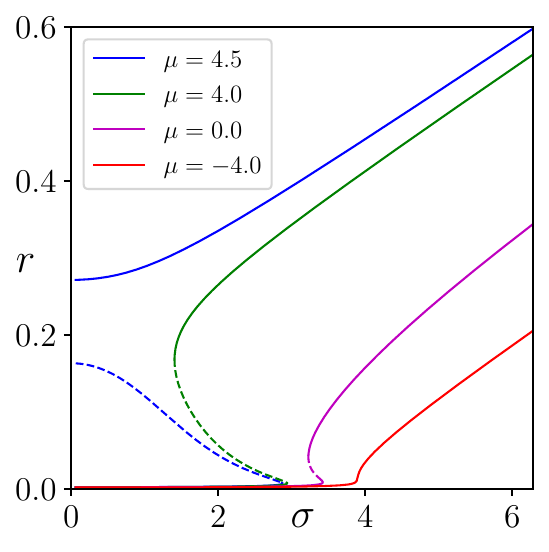}}
\put(0.1,4){\bf (c)}
\end{picture}
\begin{picture}(3.7,4)
\put(0.1,0.2){\includegraphics[height=0.235\textwidth]{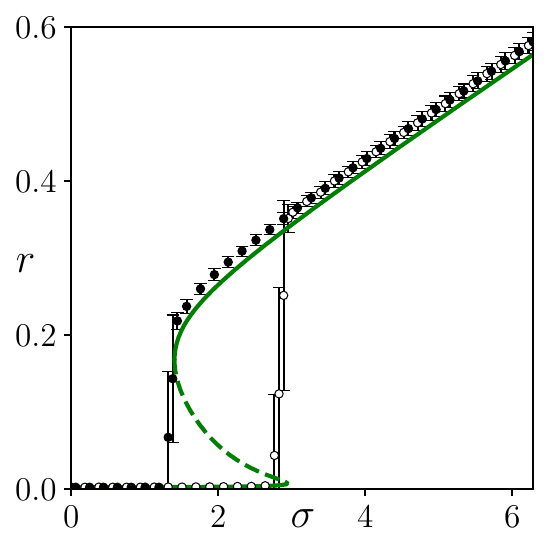}}
\put(0.2,4){\bf (d)}
\end{picture}
\caption{\textbf{Theory and simulations for Gaussian random coupling.} Shown are the self-consistent average firing rates, i.e., the neural activity, as functions of the {heterogeneity} $\sigma$ under the Gaussian white-noise approximation. We assume Gaussian random coupling Eq.\eqref{Eq:GaussianJ} and different values of mean synaptic strength $\mu$. Dashed lines correspond to unstable, and solid lines to stable solutions. The QIF neurons are excitable with $a_0=-0.5$. In (a) and (b) the noise is purely self-generated, without additional noise, i.e., $D^{add}=0$, whereas in (c) and (d) we add additional independent Gaussian white noise of strength $D^{add}=0.1$ to each neuron. In (b) and (d) we compare the theoretical stationary activity, from the mean-field dynamics, with simulations of $N=1000$ neurons and excitatory imbalance $\mu=4$. Shown are averages and one standard deviation errorbars over ten forward (open circles) and backward scans (black circles) with different realizations of $J$ (fixed $\mu$ and adjusted $\sigma$) and over $T=100$ time units for each value of $\sigma$. Without noise, the inactive state in (b) is always stable in the limit $N\to\infty$ with very small synaptic strength.}
\label{Fig:GaussTheory_vs_Sims}
\end{figure}
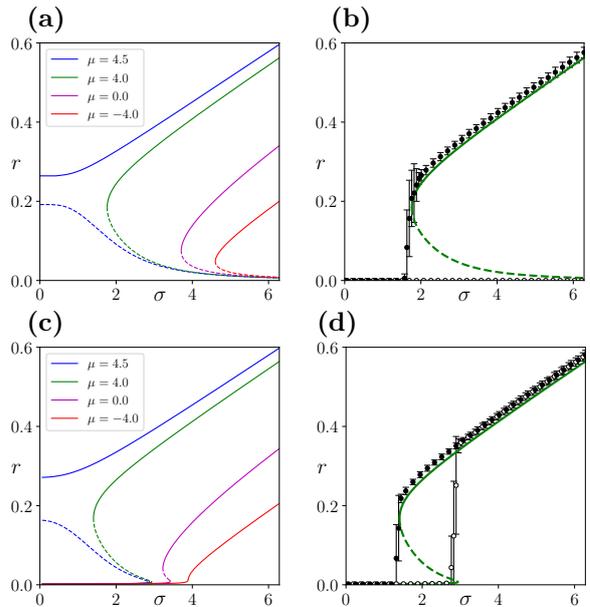
\subsection{Emergence of bistability in neural activity}
The simple theory developed above, only requires self-consistency in the mean firing rate. Other, spatio-temporal correlations in the network noise are neglected in the diffusion approximation. Nevertheless the predictions of stationary, self-sustained, asynchronous irregular firing regimes correspond quite well to the observed activities in full network simulations shown in Fig.\ref{Fig:GaussTheory_vs_Sims}b,d. For excitable neurons ($a_0<0$ as is the case for all curves in Fig.\,\ref{Fig:FoldExplorationGauss}), regions of bistability can be observed in parts of the parameter space. Drawing the self-consistent solutions, where the graphs of the input-output relation cross zero in Fig.\ref{Fig:PrePostMap}, we obtain the bifurcation diagrams in Figs.\ref{Fig:FoldExplorationGauss}a and \ref{Fig:GaussTheory_vs_Sims}. The region of bistability in parameter space, where a state of low activity, excited by weak additive noise, co-exists with a high-activity state of self-sustained firing, is defined by saddle-node bifurcations of stable and unstable self-consistent solutions. 
In Fig.\ref{Fig:FoldExplorationGauss}b we show for $a_0=-0.5$ and $\mu=0$ that the lines of the saddle node bifurcations under variation of $\sigma$ and $D^{add}$ meet in a cusp bifurcation point, where the low- and the high-activity states connect continuously but with diverging slope.

Figure \ref{Fig:GaussTheory_vs_Sims}a shows self-consistent spiking rates without additional noise, i.e., $D^{add}=0$, and for various values of imbalance $\mu$ as functions of the coupling {heterogeneity} $\sigma$. In Fig.\ref{Fig:GaussTheory_vs_Sims}c we show the solutions for the same values of $\mu$ but with additive Gaussian white noise of strength $D^{add}=0.1$.  Without the additive noise and when all neurons are at the resting potential, there is no signal in the system that could cause the neurons to fire. Furthermore, the state of zero activity is stable in the sense of $f'(r=0)<1$ for arbitrary values of $\sigma$ in the thermodynamic limit, i.e., a finite fraction of neurons must fire to carry the system across the unstable saddle into the regime of high self-sustained activity. A small amount of additive noise will create a low activity state of sparsely spiking neurons which becomes unstable at a certain value of $\sigma$ or crosses over continuously to a state of high activity. 

Figs.\ref{Fig:GaussTheory_vs_Sims}b,d compare the theoretical self-consistent firing rates, based on the Gaussian white-noise approximation (solid lines) with full network simulations of QIF neurons (data points with one standard deviation bars over ten network realizations) with excitatory imbalance $\mu=4.0$ and parameter sweeps of $\sigma$ across the bistability region. Simulations with increasing $\sigma$, starting in the low activity regime, are open circles and simulations with decreasing $\sigma$, starting in a state of high activity, are indicated by black dots. The mean activity $r$ is quite accurately predicted under the white noise approximation for excitable neurons with $a_0=-0.5$. We note that when $\mu$ is large enough, no network noise $D^{NW}=\sigma^2r/2$ from the coupling disorder is needed ($\sigma=0$) to sustain a high activity state, e.g., for $\mu=4.5$. In this purely excitatory system the bistability is mean-driven by the shift of the excitation parameter $a=a_0+\mu r$ to positive values. In contrast, for small or negative $\mu$ and low additional noise, bistability is noise-driven. Moreover, we point out, that bistability for arbitrary large values of $\sigma$ and arbitrary values of $\mu$ is only strictly true in the diffusion approximation and that the unstable self-consistent solution approaches zero asymptotically.
\section{Cauchy coupling}
\label{Sec:Cauchy}
\subsection{Cauchy white noise approximation}
Secondly, we use independent, identically distributed Cauchy random entries Eq.\,\eqref{Eq:CauchyJ} in the coupling matrix.
Because the sum of independent identically distributed Cauchy variables is also Cauchy, with additive center and scale parameter, the total synaptic input, arriving in a small enough time interval $\tau$ is also Cauchy distributed, and because of our chosen scaling with the system size in Eq.\,\eqref{Eq:CauchyJ} we obtain a system-size independent increment statistics
\begin{equation}\label{Eq:CauchyDeltaTau}
    \Delta_\tau[z^{NW}(t)]=\int_0^\tau z^{NW}(t+s)\,ds \sim \mathcal{C}(\mu{r}\tau,\sigma{r}\tau)
\end{equation}
shown in Fig.\ref{Fig:NoiseDemo}c for a single neuron.
We again approximate the network noise by white noise, but this time with Cauchy statistics, and identify the center value of the network noise with a rate-dependent shift $a^{NW}=\mu r$ in the excitation parameter, and the scale parameter with a rate-dependent strength $\Gamma^{NW}=\sigma r$ of the dynamic network noise. Note, that the integral of white Cauchy noise is Cauchy distributed with a scale parameter which grows linearly in $\tau$, whereas the standard deviation of the Gaussian white noise integral in Eq.\eqref{Eq:GaussDeltaTau} grows with the square root of $\tau$. One could say, that at small time scales, diffusion due to Gaussian white noise is much stronger than diffusion due to Cauchy white noise, while Cauchy diffusion is driven by increasingly larger jumps over longer time periods.
\subsection{Exact mean-field dynamics}
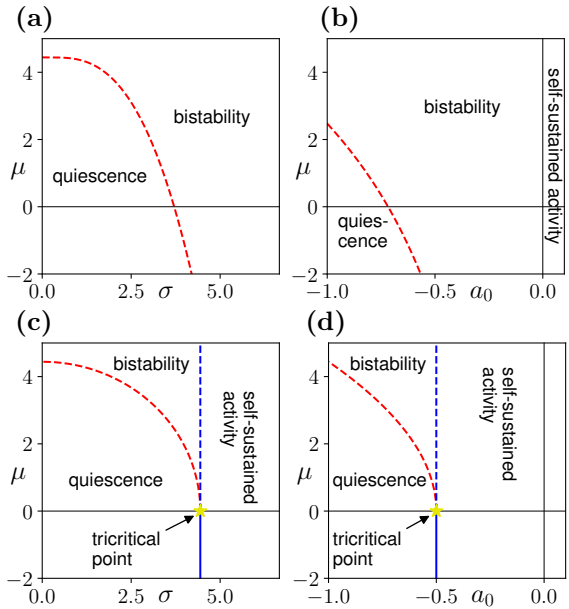
\begin{figure}[t!]
\setlength{\unitlength}{1cm}
\begin{picture}(3.6,4.0)
\put(-0.15,0.2){\includegraphics[height=0.235\textwidth]{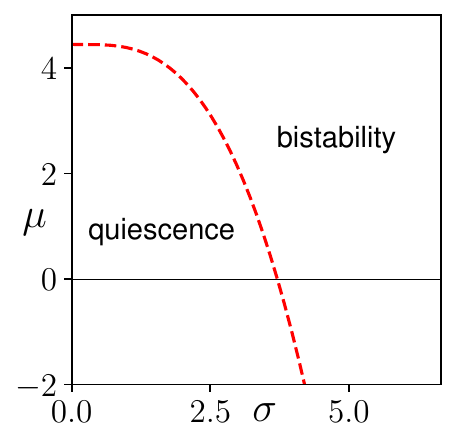}}
\put(0.1,4){\bf (a)}
\end{picture}
\begin{picture}(3.7,4.0)
\put(-0.3,0.2){\includegraphics[height=0.235\textwidth]{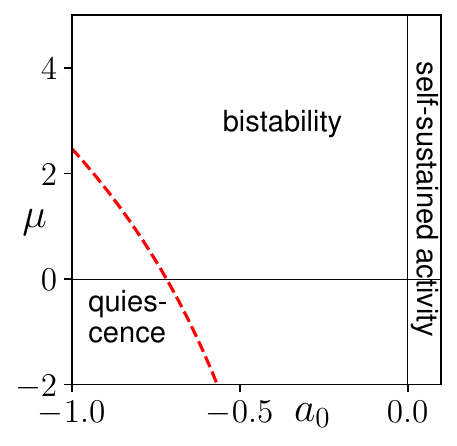}}
\put(0.2,4){\bf (b)}
\end{picture}
\begin{picture}(3.6,4.0)
\put(-0.15,0.2){\includegraphics[height=0.235\textwidth]{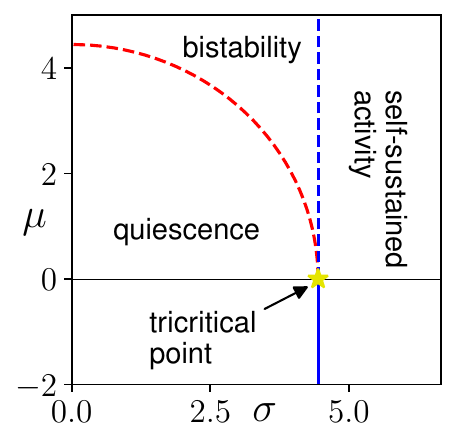}}
\put(0.1,4){\bf (c)}
\end{picture}
\begin{picture}(3.7,4.0)
\put(-0.1,0.2){\includegraphics[height=0.235\textwidth]{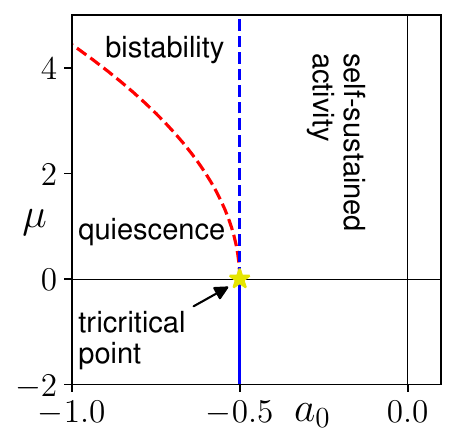}}
\put(0.2,4){\bf (d)}
\end{picture}
\caption{\textbf{Regions of qiescence, bistability and mono-stable self-sustained activity in networks without additional noise.} Shown are the saddle-node bifurcation lines (dashed red lines) where the state of self-sustained activity is created and the percolation transition (solid or dashed blue lines) where the quiescent state loses stability. Panels (a) and (b) demonstrate the Gaussiann case and (c) and (d) the case of Cauchy distributed synaptic weights, both without additional noise, i.e. $D^{add}=0$ and $\Gamma^{add}=0$. In (a) and (c) the stability regions are shown in the space of network parameters $\mu$ and $\sigma$ for excitable neurons with $a_0=-0.5$. In (b) and (d) the stability regions are shown depending on the synaptic bias $\mu$ and the excitation parameter $a_0$ with fixed heterogeneity $\sigma=2\pi\sqrt{0.5}$. When $a_0$ becomes positive, the neurons fire tonically and the quiescent state disappears. Only for heavy-tailed synaptic weight distributions the percolation transition exists. It changes from a continuous (solid blue lines) to a discontinuous transition (dashed blue lines) at a tricritical point when $\mu=0$, where it coincides with the saddle-node bifurcation. Self-organization to the tricritical point could be achieved by independent adaptation mechanisms.}
\label{Fig:Tricritical}
\end{figure}

In \cite{MonPaz15} exact dynamic equations for the ensemble mean membrane voltage ${\bar v}(t)$ and the mean firing rate ${r}(t)$ were derived for QIF neurons with quenched, Cauchy distributed excitation parameters $a\sim\mathcal{C}(a_0,\Delta)$. In \cite{CluMon24} the authors extended the analysis to also include additive white Cauchy noise of strength $\Gamma$ and showed that the mean-field dynamics depend on the sum $\Delta+\Gamma$ but otherwise have exactly the same form as in \cite{MonPaz15}. For the equivalent theta-neuron model and other phase models this equivalence of quenched Cauchy heterogeneity and white Cauchy noise was shown in \cite{ToePik20,Tanaka20}.
With explicitly rate-dependent shift $a=a_0+\mu r$ of the excitation parameter and strength $\Gamma = \Gamma^{add}+\sigma r$ of white Cauchy noise, including additive Cauchy noise of strength $\Gamma^{add}$, we modify the mean-field equations derived in \cite{CluMon24} (see also Appendix \ref{secA2}) as
\begin{eqnarray}
    \dot {r} &=& 2 {r} \bar{v}+\frac{1}{\pi}\left(\Gamma^{add}+\sigma {r}\right)\label{Eq:QIFMF1}\\
    \dot{\bar{v}} &=& \bar{v}^2 + (a_0 +\mu {r}) - \pi^2{r}^2. \label{Eq:QIFMF2} 
\end{eqnarray}
These equations can be analyzed for equilibria, where the time derivatives on the left hand sides are equal to zero. More general theories without dynamic feedback to the noise, which include transient dynamics toward the attracting Lorentzian family of the membrane voltage distributions or general $\alpha$-stable noise, have been developed in \cite{PieCesPik23,GolPer24}.\\
\begin{figure}[t!]
\setlength{\unitlength}{1cm}
\begin{picture}(3.6,4.0)
\put(-0.15,0.2){\includegraphics[height=0.235\textwidth]{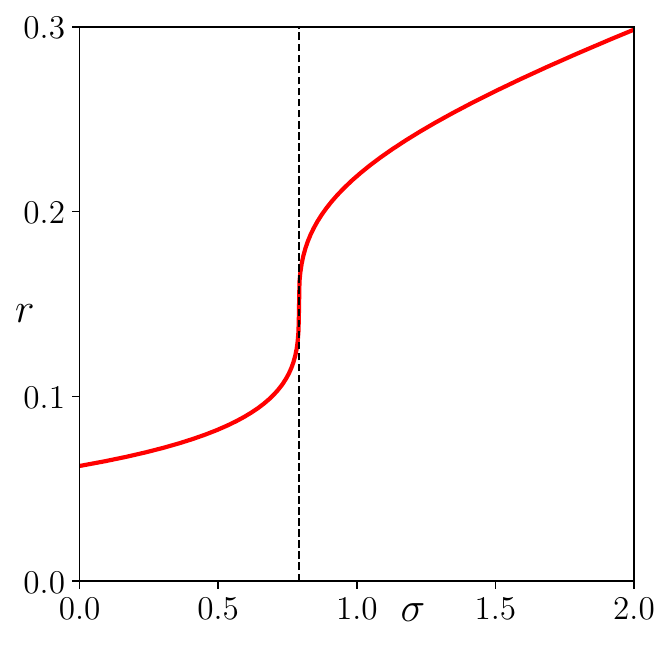}}
\put(0.1,4){\bf (a)}
\end{picture}
\begin{picture}(3.7,4.0)
\put(-0.1,0.2){\includegraphics[height=0.235\textwidth]{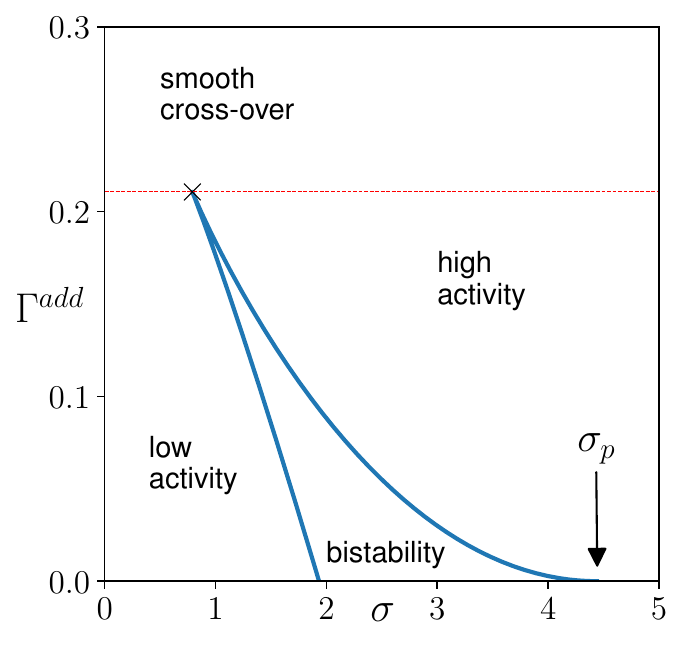}}
\put(0.2,4){\bf (b)}
\end{picture}
\caption{\textbf{Bistability region.} In case of Cauchy coupling, we obtain  $\sigma=\sigma(r)$ in Eqs.\,\eqref{Eq:Cauchy_vbar_of_r} and \eqref{Eq:Cauchy_sigma_of_r} by solving the mean-field equations \eqref{Eq:QIFMF1} and \eqref{Eq:QIFMF2} for stationary solutions and plot the resulting bifurcation diagram in (a) for $a_0=-0.5$, $\mu=4.0$ and additive Cauchy noise of strength $\Gamma^{add}=0.21$. In (b) we plot $\sigma=\sigma(r)$ against $\Gamma^{add}(r)$ from Eq.\,\eqref{Eq:Cauchy_SGMadd_of_r}. This parametric curve marks the saddle-node bifurcation of stable and unstable equilibria, i.e., the boundary of the bistability region. The additive noise strength in (a) was chosen at the height of the cusp point (dashed horizontal line in (b)), so that the low and the high activity states connect continuously with diverging slope (dashed vertical line in (a)) at a critical {heterogeneity}.}
\label{Fig:FoldExplorationCauchy}
\end{figure}
\begin{figure}[t!]
\setlength{\unitlength}{1cm}
\begin{picture}(3.6,4)
\put(-0.21,0.2){\includegraphics[height=0.235\textwidth]{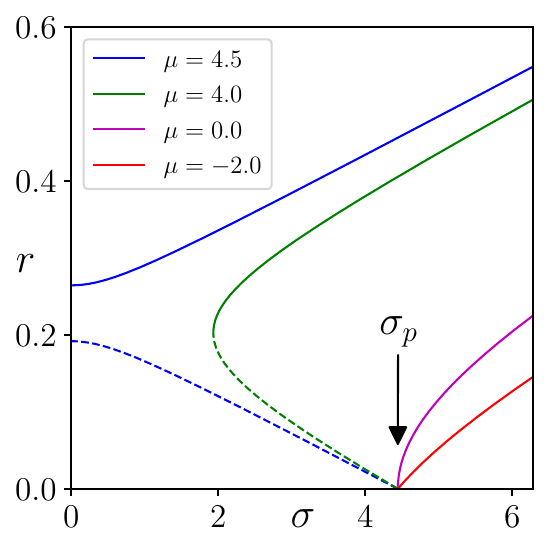}}
\put(0.1,4){\bf (a)}
\end{picture}
\begin{picture}(3.7,4)
\put(-0.1,0.2){\includegraphics[height=0.235\textwidth]{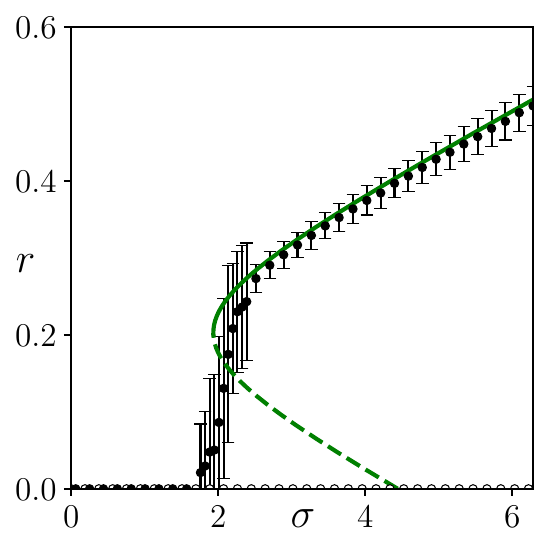}}
\put(0.2,4){\bf (b)}
\end{picture}
\begin{picture}(3.6,4)
\put(-0.21,0.2){\includegraphics[height=0.235\textwidth]{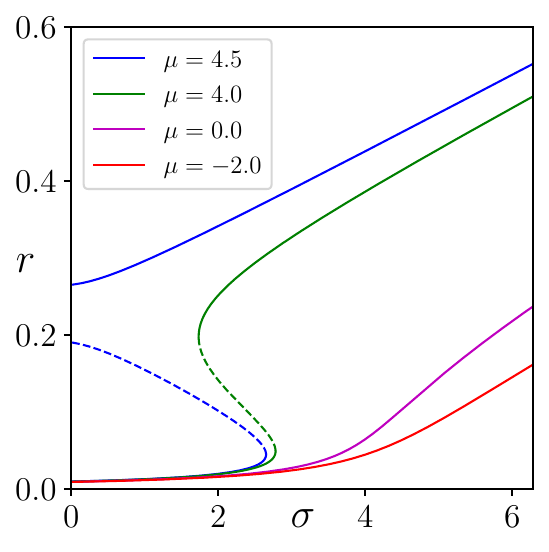}}
\put(0.1,4){\bf (c)}
\end{picture}
\begin{picture}(3.7,4)
\put(0.1,0.2){\includegraphics[height=0.235\textwidth]{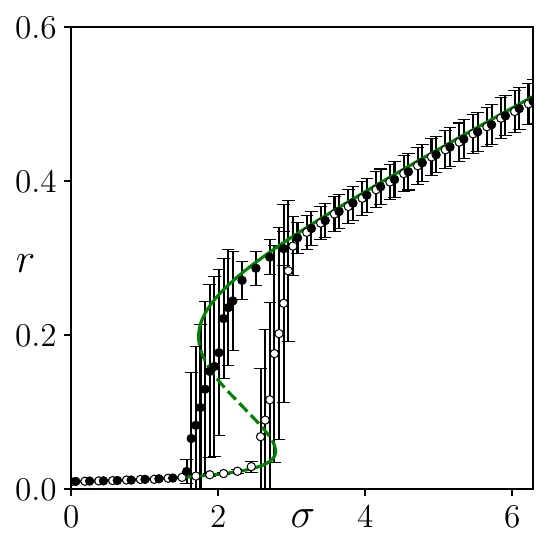}}
\put(0.2,4){\bf (d)}
\end{picture}
\caption{\textbf{Theory and simulations for Cauchy random coupling.} Shown are the self-consistent average firing rates, i.e., the neural activity, as functions of the {heterogeneity} $\sigma$ under the Cauchy white-noise approximation. We assume Cauchy random coupling Eq.\,\eqref{Eq:CauchyJ} with different center values $\mu$, i.e., excitatory-inhibitory imbalance. Dashed lines correspond to unstable, and solid lines to stable solutions. The QIF neurons are excitable with $a_0=-0.5$. In (a) and (b) the noise is purely self-generated, without additional noise, i.e., $\Gamma^{add}=0$. In this case all solutions connect to the $\sigma$-axis at the percolation threshold $\sigma_p=2\pi\sqrt{-a_0}$. In (c) and (d) we add additional independent white Cauchy noise of strength $\Gamma^{add}=0.04$ to each neuron. In (b) and (d) we compare the theoretical stationary activity, from the mean-field dynamics, with simulations of $N=1000$ neurons and excitatory imbalance $\mu=4$. Shown are averages and one standard deviation errorbars over ten forward (open circles) and backward scans (black markers) with different realizations of $J$ (fixed $\mu$ and adjusted $\sigma$) and over $T=100$ time units for each value of $\sigma$. Without noise, the inactive state in the network simulations in (b) is always stable against local perturbations of the membrane voltages from the resting value, but it is unstable against a vanishing fraction of input spikes above the percolation threshold at $\sigma_p$ (not shown).}
\label{Fig:CauchyTheory_vs_Sims}
\end{figure}
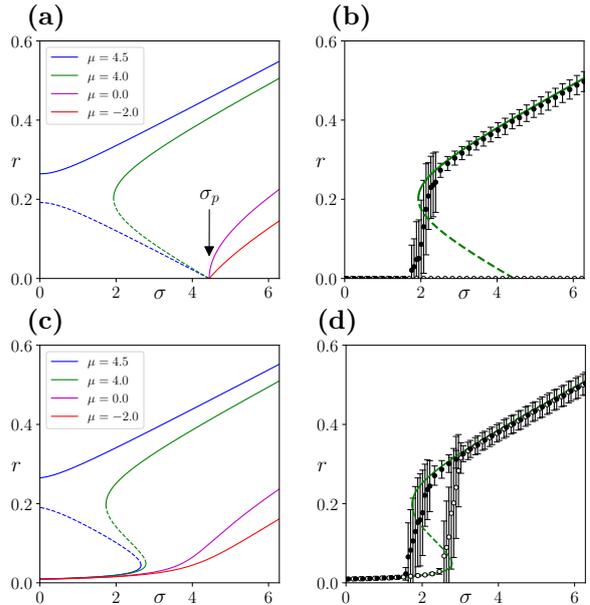
First, for $\Gamma^{add}=0$ without additional noise, a trivial pair of stable and unstable equilibria exists for excitable QIF neurons when all neurons are at the resting potential $v_\ell=\bar{v}=-\sqrt{-a_0}$ or at the saddle $v_\ell=\bar{v}=\sqrt{-a_0}$, both with zero network activity. For ${r}\ne 0$ the equilibrium mean potential from setting the left hand side in Eq.\,\eqref{Eq:QIFMF1} to zero is 
\begin{equation}
    \bar{v} = -\frac{\sigma}{2\pi}
\end{equation}
independent of $r$, $a_0$ and $\mu$.
Putting this solution into Eq.\,\eqref{Eq:QIFMF2} and setting the left hand side to zero, the firing rate $r$ is found as a non-negative solution of
\begin{equation}
    \frac{\sigma^2}{4\pi^4} = {r}^2-\frac{\mu}{\pi^2}{r}-\frac{a_0}{\pi^2},
\end{equation}
which describes hyperbolas in the $(\sigma,{r})$ plane. Only the parts of the hyperbolas in the positive quadrant $\sigma,{r}\ge 0$ correspond to physical, self-consistent equilibria. 
For excitable neurons $a_0<0$, and when ${r}\to 0$, all solutions connect to the $\sigma$ axis at the critical point
\begin{eqnarray}
    \sigma_{p} = 2\pi\sqrt{-a_0}.
\end{eqnarray}
At this critical {heterogeneity} there is a directed percolation transition \cite{KusOga20}. In particular, for $\mu=0$ the percolation transition is tricritical \cite{Lueb06,PiuMar23}, i.e., through a point in parameter space where three phases meet, quiescence, monostable and bistable self-sustained activity (see Fig.\ref{Fig:Tricritical}). The activity scales as a square root ($\beta=1/2$) with the distance $(\sigma-\sigma_p)$ to the transition point
\begin{equation}\label{Eq:beta05}
    r = \sqrt{\sigma^2-4\pi^2a_0}=\sqrt{\sigma-\sigma_p}\sqrt{\sigma+\sigma_p}
\end{equation}
shown as the purple curve in Fig.\ref{Fig:CauchyTheory_vs_Sims}a. 

The directed percolation nature of this transition is demonstrated analytically by calculating the expected number of synapses for each neuron with a strength larger than $2\sqrt{-a_0}$, i.e., the difference between the resting potential and the saddle. This expected number is equivalent to the basic reproduction number $R_0$ in epidemic theory, when one spike (infection) in a presynaptic neuron elicits (infects) on average one spike in a postsynaptic neuron.
\begin{eqnarray}
    R_0 &=& \lim_{N\to\infty} N\cdot \textrm{Prob}\left(J_{\ell m}\ge 2\sqrt{-a_0}\right) \nonumber \\
    &=&\lim_{N\to\infty} N\int_{2\sqrt{-a_0}}^\infty \frac{1}{\pi} \frac{\sigma/N}{(J-\mu/N)^2+\sigma^2/N^2}\,dJ\nonumber \\
    &=& \lim_{N\to\infty} \frac{N}{\pi}\left[\frac{\pi}{2}-\arctan\left(\frac{N} {\sigma}\left(2\sqrt{-a_0}-\frac{\mu}{N}\right)\right)\right] \nonumber \\
    &=& \frac{\sigma}{2\pi\sqrt{-a_0}}=1 \quad\textrm{for } \sigma=\sigma_p 
\end{eqnarray}
where we have used $\tfrac{\pi}{2}-\arctan\left(\tfrac{1}{x}\right)=\arctan(x)\approx x$ for $x\to 0$. The transition only exists for excitable neurons, i.e., when $a_0<0$. 
\\ 
Note, that any distribution with support in the positive numbers leads to a percolation transition  if the scale parameter is large enough or if the excitation parameter $a_0\to 0^-$ is small enough. For instance, with a binary distribution $p(J)=\frac{1}{2}\left[\delta(J-\sigma/\sqrt{N})+\delta(J+\sigma/\sqrt{N})\right]$ activity percolates discontinuously, when $\sigma>\sigma_p=2\sqrt{-a_0N}$. Similarly for a Gaussian synaptic weight distribution \eqref{Eq:GaussianJ} asymptotic analysis also shows $\sigma_p \sim \sqrt{-a_0 N}$. The existence of a percolation threshold is the direct consequence of delta-pulse coupling and not predicted in the diffusion approximation.

Interestingly, a positive feedback of the network activity on the excitation parameter $a^{NW}=\mu{r}>0$ makes the transition in the Cauchy network also discontinuous. In this case, as for excitable neurons with Gaussian coupling, the activity in the network is bistable. The high activity state and the unstable equilibrium collide in a saddle node bifurcation, which for Cauchy random coupling and without additive noise is located at 
\begin{equation}
    \sigma_b = \sqrt{-\mu^2-4\pi^2 a_0 },\qquad {r}_b=\frac{\mu}{2\pi^2}.
\end{equation}
Or the state of high activity connects to the ${r}$-axis at $\sigma=0$, ${r}=(\mu+\sqrt{\mu^2+4\pi^2a_0})/2\pi^2$, when $\mu^2+4\pi^2a_0>0$ (See Fig.\ref{Fig:CauchyTheory_vs_Sims}a for $\mu=4.5$). The lower branch of the hyperbola is always an unstable saddle, while the upper branch is always stable.

When we include additive Cauchy noise of constant strength $\Gamma^{add}>0$ the bifurcation lines are smoothed out and the percolation transition disappears. In this case we can draw the bifurcation diagram parametrically $(\sigma({r}),{r})$ as a function of ${r}>0$ (Fig.\ref{Fig:CauchyTheory_vs_Sims}c). The second mean-field equation Eq.\,\eqref{Eq:QIFMF2} with $\dot{\bar{v}}=0$ defines the square of the stationary mean membrane potential
\begin{equation}\label{Eq:Cauchy_vbar_of_r}
    \bar{v}^2 = \pi^2 {r}^2 - \mu {r} - a_0
\end{equation}
and the first mean-field equation Eq.\,\eqref{Eq:QIFMF1} with $\dot{r}=0$ defines the coupling {heterogeneity}
\begin{equation}\label{Eq:Cauchy_sigma_of_r}
    \sigma({r}) = -\frac{\Gamma^{add}}{{r}} + 2\pi\sqrt{\bar{v}^2}.
\end{equation}
The saddle-node bifurcations of stable-unstable equilibria are located where $\partial\sigma/\partial r=0$, at $\sigma(r)$ given by Eq.\,\eqref{Eq:Cauchy_sigma_of_r} and with additive noise strength
\begin{equation} \label{Eq:Cauchy_SGMadd_of_r}
    \Gamma^{add}(r) = \frac{2\pi^2 r - \mu}{\sqrt{\pi^2r^2 - \mu r - a_0}}-\pi r^2.
\end{equation}
The lines of saddle-node bifurcations form the region of bistability in the $(\sigma,\Gamma^{add})$ parameter space as shown in Fig.\,\ref{Fig:FoldExplorationCauchy}b.
\subsection{Continuous and discontinuous transitions to self-sustained activity}
We perform the simulations in the same way as for the Gaussian coupling case. However, we use additive, white Cauchy noise instead of Gaussian white noise, i.e., we add independent random increments $\Delta w \sim \Delta t\,\Gamma^{add} \mathcal{C}(0,1)$ to each membrane potential in each time step, in addition to the synaptic weights of all firing presynaptic neurons.
\\ \\
For low {heterogeneity}, and when we do not include additive noise, the residual activity is very low or zero. Above the critical {heterogeneity} $\sigma>\sigma_{p}$ the activity increases and all neurons participate in self-sustained irregular spiking. Depending on the excitatory-inhibitory imbalance $\mu$, this bifurcation is found to be transcritical continuous for $\mu<0$ and transcritical discontinuous for $\mu>0$, in both cases with critical exponent $\beta=1$. 
As already mentioned above, only in the special case $\mu=0$ the critical exponent of the continuous bifurcation is $\beta=1/2$ (see Eq.\,\eqref{Eq:beta05}). While it is necessary to tune two parameters, e.g., $\mu$ and $\sigma$ of the synaptic strength distribution, to poise the system at this tricritical point (Fig.\ref{Fig:Tricritical}), the two adaptation mechanisms for $\mu\to 0$ and $\sigma\to\sigma_p$ could be independent. A slow activity dependent build-up and consumption of a resource can tune the synaptic strengths or the excitation parameter $a_0$ to the percolation threshold \cite{LevHer07}, while exact balance could be the self-organized result of synaptic plasticity \cite{LuzSha12,Fro15}.

When additive noise is included, the quiescent state becomes a state of low firing rate with occasional noise induced spikes, the region of bistability becomes smaller, and continuous phase transitions turn into smooth cross-overs (see Figs.\,\ref{Fig:FoldExplorationCauchy} and \ref{Fig:CauchyTheory_vs_Sims}c). 
\\ \\
In \cite{KusOga20} it was reported, analytically for a neuron model of coupled nonlinear maps, and numerically for excitable integrate-and-fire neurons, that Cauchy random coupling leads to a continuous phase transition of directed percolation type which is not present for Gaussian random coupling in the infinite system size limit under appropriate scaling. 
The authors point out that a continuous phase transition is necessary for the mechanism of self-organized criticality to poise the network at the critical point which would explain the power-law scaling of activity reported in biological experiments. 

In our analytical treatment of QIF neurons subject to independent white Cauchy noise we derive the full bifurcation diagram. We find that the directed percolation transition remains continuous for slightly more inhibitory synapses ($\mu<0$) but the activity scales with the critical parameter $\beta=1$.

The emergence of bistability through a change from a continuous to a discontinuous transcritical bifurcation has recently been demonstrated in the mean-field description of a stochastic Wilson-Cowan model in networks with excitatory-inhibitory imbalance \cite{PiuMar23}. In the studies \cite{PiuMar23,KusOga20} a generic hyperbolic tangent and a step function were used as nonlinear activation functions and the impact of dynamic network noise was not considered.

Unsuprisingly, the transcritical bifurcation disappears for any amount of additive noise and is replaced by a fold-bifurcation with qualitatively similar behavior to the case of Gaussian random coupling, where a state of low, randomly excited activity co-exists with a state of high, self-sustained irregular firing. In both cases the cusp bifurcation point would be another candidate for a self-organized critical state with critical exponent $\beta=1/3$ (see Fig.\,\ref{Fig:FoldExplorationCauchy}a). However, this requires a more involved coordinated adaptation of two parameters to the critical point. 
\section{Summary}
\label{sec:conclusion}
We have used the white-noise approximation for synaptic inputs to self-consistently find steady states in networks of randomly coupled spiking quadratic integrate-and-fire neurons. 
Our main focus was to compare the steady states of the firing rate for Cauchy and for Gaussian coupling statistics, corresponding to synaptic weight distributions with or without heavy tails. 
Qualitatively, both coupling statistics can give rise to a bistability of the firing rate, i.e., a coexistence of a state of low randomly excited activity and of a state of high self-sustained activity. Furthermore, in both scenarios, additional external noise makes the range of bistability smaller, and eventually disappear in a cusp point. The simultaneous tuning of two system parameters is necessary to poise a network of spiking neurons at this critical cusp point. 

However, there are also qualitative differences between the two cases. With Gaussian coupling statistics, the region of bistability extends to arbitrarily large coupling {heterogeneities} $\sigma$ and both positive and negative values of excitatory-inhibitory imbalance $\mu$. In contrast, in the Cauchy case, bistability is only observed for dominating recurrent excitation ($\mu>0$) and below a critical value $\sigma\le \sigma_p$ of the coupling {heterogeneity}. 

Only in the case of heavy-tailed synaptic weights, inhibitory imbalance ($\mu\le 0$) and no additional noise ($\Gamma^{add}=0$), a continuous directed percolation transition for the firing rate emerges when either the coupling {heterogeneity} $\sigma$ or the excitation parameter $a_0$ is varied.
If the network is furthermore exactly balanced ($\mu=0$) this transition has an order parameter critical exponent $\beta=1/2$, as was demonstrated for a time-discrete neuron model analytically, and numerically for integrate-and-fire neurons in \cite{KusOga20}. In networks with Gaussian (light-tailed) synaptic weight distributions bistability is prevalent and a continuous phase transition from quiescence to self-sustained activity is generally not observed.
%
%

\backmatter

\bmhead{Acknowledgements}
RT and BL acknowledge funding from DFG grants LI 1046/10-1 and LI 1046/6-2.
\section*{Declarations}
\subsection*{Competing interests}
The authors declare no competing interests related to this work, financial or otherwise.
\subsection*{Author contribution}
RT contributed the mathematical analysis and numerical simulations of the models. CZ initiated the project and also contributed to the mathematical modeling and to numerical simulations. WC's insights into dynamic mean-field theory were valuable contributions to the project. BL directed the research from a neuroscience perspective and wrote the text together with RT.

\begin{appendices}

\section{}\label{secA1}
{\bf Parametric self-consistent solutions, Gaussian case}\\ \\
A self-consistency equation like \eqref{Eq:RateConsistency} defines a hypersurface in the combined space of system parameters and observables, and thus implicitly the self-consistent firing rate for any given set of system parameters. We cannot simply solve \eqref{Eq:RateConsistency} for the firing rate in terms of elementary functions. The direct way would be to use numeric root finding to solve this as a one dimensional fixed point problem. However, it is often possible to solve an equation for a different variable and then plot this variable as a function of the rate. Or we define a proxy variable, say $s$, which combines the firing rate and other system parameters and eliminates the rate on the right hand side of \eqref{Eq:RateConsistency}. We can then solve two equations to obtain a parametric solution e.g. $(\sigma(s),r(s))$. We have to calculate the values of the convergent power series $G(s)$ in Eq.\,\eqref{Eq:QIF_Gs} or use its asymptotics when $|s|\gg 1$, but no root finding or numerical quadrature of integrals is necessary.
\\ \\
We define an auxilliary variable $t$, use the definition of $s$,  and express $r$ in terms of $t$ and $G(s)$ given as power a series in Eq.\,\eqref{Eq:QIF_Gs} 
\begin{eqnarray}
        t &=& \left(\frac{3D}{2}\right)^\frac{1}{3}
        = \left(\frac{3(D^{add}+\tfrac{\sigma^2}{2} r)}{2}\right)^\frac{1}{3},
        \label{Eq:ApdxA_t}\\
        s &=& -a\left(\frac{12}{D^2}\right)^\frac{1}{3} = -\frac{3a}{t^2}=-3\frac{a_0+\mu r}{t^2}, \label{Eq:ApdxA_s}\\
        r &=& \frac{t}{G(s)}. \label{Eq:ApdxA_r}
\end{eqnarray}
Substituting $r$ with \eqref{Eq:ApdxA_r} into \eqref{Eq:ApdxA_t} and \eqref{Eq:ApdxA_s}, and taking $t$ in \eqref{Eq:ApdxA_t} to the third power we arrive at the two algebraic equations
\begin{eqnarray}
    st^2G + 3a_0G +3\mu t &=& 0 \label{Eq:ApdxA_F1}\\
    2t^3G - 3 D^{add}G-\frac{3\sigma^2}{2}t&=&0. \label{Eq:ApdxA_F2}
\end{eqnarray}
Solving the quadratic equation \eqref{Eq:ApdxA_F1} for $t$ gives at most two branches of parametric solutions. The second equation can then be used to express $D^{add}$ or $\sigma$ as a function of $s$. Or we solve the depressed cubic equation \eqref{Eq:ApdxA_F2} for $t$, with at least one and at most three branches, and use the first equation to express $a_0$ or $\mu$ as a function of $s$. Solutions must be real valued and $r, \sigma^2$ and $t$ must be non-negative to be physically meaningful. The free parameter $s$ can be positive, negative or change sign depending on the values of $a_0$ and $\mu r$.
\\ \\
The saddle node bifurcation of the upper state of self-sustained activity and the unstable saddle occur under the additional condition $\tfrac{d}{ds}\sigma^2 = 0$. Dividing \eqref{Eq:ApdxA_F2} by $t$ and differentiating both equations \eqref{Eq:ApdxA_F1} and \eqref{Eq:ApdxA_F2} with respect to $s$ yields
\begin{eqnarray}
    t^2G+2st'tG+st^2G' + 3a_0 G' + 3\mu t' &=& 0 \label{Eq:ApdxA_dF1}\\
    4t'tG + 2t^2G' + 3D^{add}\left(\frac{G'}{t}-\frac{Gt'}{t^2}\right)&=& 0\quad \label{Eq:ApdxA_dF2}
\end{eqnarray}
where
\begin{equation}
    G'(s) = \sqrt{\frac{\pi}{3}} \sum_{n=0}^\infty \frac{s^n}{n!}\Gamma\left(\frac{2n+3}{6}\right). \label{Eq:ApdxA_dG}
\end{equation}
Eqs. \eqref{Eq:ApdxA_F1}-\eqref{Eq:ApdxA_dF2} are linear in $\sigma^2, D^{add}, a_0, \mu, t'$ and nonlinear in $t$. Using \eqref{Eq:ApdxA_r}, a solution of these equations uniquely determines the self-consistent rate $r(s)$ at the saddle-node bifurcation point.
\section{}\label{secA2}
{\bf Mean-field equations, Cauchy case}\\ \\
The exact low-dimensional mean-field equations for QIF neurons with independent additive Cauchy noise on the invariant family of Lorentzian membrane voltage distributions have been derived in \cite{CluMon24} using the formalism of the fractional Fokker-Planck equation. More general mean-field equations with different noise statistics or away from the Lorentzian distribution family have been derived in \cite{PieCesPik23,GolPer24}. Here we follow the derivation without Cauchy-noise from the non-fractional continuity equation and then motivate how the addition of Cauchy noise changes the mean-field dynamics, skipping the mathematically rigorous derivation.
\\
We first show that the scale parameter $\pi r$ of the family of Cauchy (Lorentzian) distributions
\begin{equation}
    p(v) = \frac{1}{\pi}\frac{\pi{r}}{(v-{\bar v})^2+\pi^2{r}^2}
\end{equation}
is indeed proportional to the firing rate $r$ under the deterministic QIF dynamics $\dot{v}=v^2+a_0$. The firing rate equals $r=\lim_{v\to\infty} \dot{v}p(v)$, i.e. the probability current at the firing threshold $v\to\infty$. We check that
\begin{equation}
    r = \lim_{v\to\infty} (v^2+a_0) \frac{1}{\pi}\frac{\pi r}{(v-\bar{v})^2+\pi^2 r^2}
\end{equation}
follows from L'Hospital's rule.
\\
Next, we show that the family of Cauchy distributions 
is invariant under the continuity equation
\begin{equation}\label{Eq:RiccatiContinuity}
    \partial_t p = -\partial_v\left((v^2+a_0)p\right)
\end{equation}
for membrane voltages following the deterministic QIF dynamics if and only if the parameters $\bar{v}=\bar{v}(t)$ and $r=r(t)$ follow the mean-field differential equations
\begin{eqnarray}
    \dot r &=& 2r\bar{v} \label{Eq:RiccatiMF0_r}\\
    \dot{\bar{v}} &=& \bar{v}^2 + a_0 - \pi^2 r^2. \label{Eq:RiccatiMF0_v}
\end{eqnarray}
For this, we perform the derivatives in Eq.\,\eqref{Eq:RiccatiContinuity}
\begin{eqnarray}\label{Eq:RicconLeft}
    \partial_t p &=& \dot r \partial_r p + \dot{\bar{v}}\partial_{\bar{v}}p \\
    &=&\dot r \left[\left(\frac{1}{(v-\bar{v})^2+\pi^2r^2}\right)-\frac{2\pi^2r^2}{\left((v-\bar{v})^2+\pi^2r^2\right)^2}\right]\nonumber \\
    &+& \dot{\bar{v}} \frac{ 2 (v-\bar{v}) r}{\left((v-\bar{v})^2+\pi^2r^2\right)^2} \nonumber
\end{eqnarray}
and
\begin{eqnarray}\label{Eq:RicconRight}
    &&-\partial_v\left((v^2+a_0)p\right) = -2vp - (v^2+a_0)\partial_v p \\
    &=& - \frac{2vr}{(v-\bar{v})^2+\pi^2r^2} + \frac{(v^2+a_0)2(v-\bar{v})r}{\left((v-\bar{v})^2+\pi^2r^2\right)^2}\nonumber.\quad\qquad
\end{eqnarray}
We multiply both sides of \eqref{Eq:RiccatiContinuity}, i.e., \eqref{Eq:RicconLeft} and \eqref{Eq:RicconRight} with the nonzero denominator $\left((v-\bar{v})^2+\pi^2r^2\right)^2$, bring all terms to the right hand side, sort by powers of $v$ and obtain
%
\begin{eqnarray}
    0 =&v^2&  \left(2r\bar{v}-\dot r\right) \nonumber \\
       +&v&\left(2\bar{v}\dot r - 2\dot{\bar{v}} r + 2ra_0-2r\bar{v}^2-2\pi^2r^3\right)  \qquad\quad \\
       +&&\left(2r\bar{v}\dot{\bar{v}} + \pi^2r^2\dot r-\dot r \bar{v}^2 - 2ra_0\bar{v}\right). \nonumber
\end{eqnarray}
The right hand side is zero independently of the values $v$ if and only if the brackets are zero, which is exactly the case when \eqref{Eq:RiccatiMF0_r} and \eqref{Eq:RiccatiMF0_v} hold. The family of Cauchy distributions is invariant under the deterministic QIF dynamics if and only if the parameters follow the mean-field dynamics Eqs.\eqref{Eq:RiccatiMF0_r} and \eqref{Eq:RiccatiMF0_v}.
\\ \\
Adding biased, white Cauchy noise $\Delta w\sim \mu r \Delta t+ (\Gamma^{add}+\sigma r) \Delta t \mathcal{C}(0,1)$ in each time step, and observing the additivity of the center and scale parameters for sums of independent Cauchy distributed random variables $v$ and $\Delta w$, one can argue \cite{Tanaka20}, that the time derivatives of the center and the scale parameter must be modified as
\begin{eqnarray}
    \dot r &=& 2r\bar{v} + \frac{1}{\pi}\left(\Gamma^{add}+\sigma r\right) \label{Eq:RiccatiMF1_r}\\
    \dot{\bar{v}} &=& \bar{v}^2 + (a_0 + \mu r) - \pi^2 r^2. \label{Eq:RiccatiMF1_v}
\end{eqnarray}
These are the mean-field equations used in Sec.\,\ref{Sec:Cauchy}.
\end{appendices}

\bibliographystyle{apsrev4-2}


\end{document}